# Physical origin of superconductivity in $EuFe_2As_{1.4}P_{0.6}$ and $EuFe_2As_2$: pressure-induce valence change of europium

Liling Sun[1]*, Jing Guo[1], Genfu Chen[2] Xianhui Chen[3], Xiaoli Dong[1], Wei Lu[1], Chao Zhang[1], Zheng Jiang[4], Bo Zou[4], Yuying Huang[4], Qi Wu[1], Xi Dai[1] and Zhongxian Zhao[1]*

[1]Institute of Physics and Beijing National Laboratory for Condensed Matter Physics, Chinese Academy of Sciences, Beijing 100190, China;
[2]Department of Physics, Renmin University of China, Beijing 100872, China
[3] Hefei National Laboratory for Physical Science at Microscale and Department of Physics, University of Science and Technology of China, Hefei, Anhui 230026, China
[4]Shanghai Synchrotron Radiation Facilities, Shanghai Institute of Applied Physics, Chinese Academy of Sciences，Shanghai 201204, China

Superconductivity can be realized in Eu-containing pnictides by application of chemical (internal) and physical (external) pressure, the intrinsic physical mechanism of which attracts much attention in physics community. Here we present the experimental evidence for the valence change of europium in compounds of $EuFe_2As_{1.4}P_{0.6}$ exposed to ambient pressure and $EuFe_2As_2$ to high pressure by x-ray absorption measurements on $L_3$-Eu edge. We find that the absorption spectrum of $EuFe_2As_{1.4}P_{0.6}$ at ambient pressure shows clear spectra weight transfer from a divalent to a trivalent state. Furthermore, application of pressure on $EuFe_2As_2$ using a diamond anvil cell shows a similar behavior of valence transition as $EuFe_2As_{1.4}P_{0.6}$. These findings are the first observation of superconductivity mechanized by valence change in pnictides superconductors and elucidate the intrinsic physical origin of superconductivity in $EuFe_2As_{1.4}P_{0.6}$ and compressed $EuFe_2As_2$.

Corresponding authors:
llsun@aphy.iphy.ac.cn
zhxzhao@aphy.iphy.ac.cn

The discovery of Fe-based superconductor [1] breaks a new path for understanding of high temperature superconducting mechanism which has been enigmatic for more than two decades. Experimental studies show that appropriate charge carrier doping or compressing of the parent compounds can suppress the spin density wave （SDW）transition originated from the long range antiferromagnetic (AFM) order of the Fe moments at low temperature and expedite appearance of superconducting state [2-10]. Among the parent compounds in iron arsenide family, $EuFe_2As_2$ is a unique member because it exhibits a strong additional moments of $Eu^{2+}$ ions ordered antiferromagnetically at ~20 K in addition to AFM order of the Fe moments [11-14]. It is reported that the superconducting transition temperature of 30 K at 2.6-2.8 GPa has been observed in $EuFe_2As_2$ when physical (external) pressure is applied and the SDW is suppressed [15-16]. By replacing Eu by alkali metals, superconducting transition has been achieved at 31K for $Eu_{0.5}K_{0.5}Fe_2As_2$ and 34.5K for $Eu_{0.7}Na_{0.3}Fe_2As_2$ [17-18]. Remarkably, substitution phosphorus for arsenic in $EuFe_2As_2$ exhibits superconductivity at 26 K at ambient pressure [19]. It is known that the isovalence substitution induces zero charge carrier to the system, but the superconducting transition has been found in $EuFe_2As_{1.4}P_{0.6}$ which is deemed to be attributed to the chemical pressure induced by substitution of phosphorus (phosphorus has smaller ionic radius than arsenic). The experimental evidence obtained so far implies that either chemical or physical pressure can turn the parent compound to superconductivity. However, what driving superconducting mechanism behind the superconductivity is in compressed $EuFe_2As_2$ (Tc=26 K for P-doped sample and Tc=

30 K for parent sample pressurized to 2.6 GPa) remains unclear.

In the studies of possible connection to physical mechanism of superconductivity in $EuFe_2As_{1.4}P_{0.6}$ and compressed $EuFe_2As_2$, we note that the volume of the divalent metal Eu with a $4f^7$ electron shell is larger than that of the trivalent Eu with a $4f^6$ electron shell. It is anticipated that the valence of Eu ions in $EuFe_2As_2$ may transfer from the divalent to the trivalent upon increasing pressure. X-ray absorption spectroscopy (XAS) is one of the most important and useful techniques for determination of valence transition, which has been widely applied in physics and chemistry. In this report, we demonstrate ambient-pressure XAS studies on Eu ions for single crystals $EuFe_2As_2$, $EuFe_2As_{1.4}P_{0.6}$ and $EuFe_{1.715}Co_{0.285}As_2$. A clear valence transition from $Eu^{2+}$ to $Eu^{3+}$ is found only in $EuFe_2As_{1.4}P_{0.6}$ sample, revealing an internal pressure role for the valence transition. We also made high-pressure XAS studies on Eu ions for single crystal $EuFe_2As_2$ which exhibits superconductivity at 2.6-2.8 GPa [15-16]. Application of external pressure on *ab* plane of $EuFe_2As_2$ sample, the valence change of $Eu^{2+}$ ions is observed upon uploading beginning at 2.1 GPa. With further increasing pressure, the mean valence of the Eu ions increases and saturates at pressure above 9.8 GPa. For comparison, we also performed XAS measurements on single crystal $EuFe_{1.715}Co_{0.285}As_2$ and did not see any change in valence for the divalent Eu ions in the sample. The fact that the increment of Eu's valence state, which is induced by internal or external pressure, gives an evidence for charge transfer from the Eu ions to the $FeA_S$ layer, and implies that pressure-induced valence transition has a strong connection with superconducting mechanism of

$EuFe_2As_{1.4}P_{0.6}$ and compressed $EuFe_2As_2$.

High quality single crystals with nominal composition $EuFe_2As_2$ and $EuFe_2As_{1.4}P_{0.6}$ were prepared by solid reaction methods. First, the polycrystalline $EuFe_2As_2$ and $EuFe_2As_{1.35}P_{0.65}$ were prepared in evacuated quartz tubes at 900 ℃ for 50 h and then the resulting materials were ground thoroughly, pressed into pellets, and loaded into $Al_2O_3$ crucibles. Then, the crucibles were sealed in evacuated quartz tubes and were heated to 1190 ℃, held at this temperature for 24 h, and finally cooled to 1100 ℃ over 100 h. Many plate-like $EuFe_2As_2$ and $EuFe_2As_{1.4}P_{0.6}$ crystals were obtained. Ambient-pressure and high-pressure XAS experiments were performed at Shanghai Synchrotron Radiation Facility (SSRF). High pressure was created using a diamond anvil cell. To maintain the sample in a hydrostatic pressure environment, Daphne 7373 was used as pressure medium in the high-pressure measurements. Pressure was determined by ruby fluorescence [20].

Figure 1 shows x-ray diffraction patterns of $EuFe_2As_2$, $EuFe_2As_{1.4}P_{0.6}$ samples at room temperature. Only (00l) reflections can be detected from 2θ scan, indicating that the samples investigated are single crystals. Calculations for c axis yields c=1.207(7) nm, 1.188(7) nm for $EuFe_2As_2$, and $EuFe_2As_{1.4}P_{0.6}$ respectively, which are closed to the c value reported [19]. Obviously, c value of P-doped sample is reasonably smaller than that of $EuFe_2As_2$, demonstrating that the lattice was indeed reduced by phosphorous doping. In Fig.2, we present temperature dependence of electrical resistance, magnetization and ac susceptibility data of the two samples. For the parent compound, the resistance exhibits an anomaly at ~173 K which has been confirmed to be associated with SDW and structure transition [13,16]. The kink at ~ 19 K is related to magnetic ordering of localized $Eu^{2+}$ moment. With phosphorus doping, the SDW transition is completely suppressed and the zero resistance has been observed (inset of Fig.2(a)). Plunge of resistance at 19 K in $EuFe_2As_{1.4}P_{0.6}$ sample is a sign of superconducting transition. Close inspection of resistance data below 19 K, a kink is found at ~7 K that should be attributed to $Eu^{2+}$ magnetic ordering. The dc magnetization data show that the increasing tendency of magnetization of these two samples is depressed significantly below 20 K. As reported in Ref [19], no diamagnetization is observed from the P-doped sample. In order to further indentify coexistence of the ferromangnetic ordering and superconducting state in $EuFe_2As_{1.4}P_{0.6}$, we performed ac susceptibility for $EuFe_2As_2$, $EuFe_2As_{1.4}P_{0.6}$ and $EuFe_{1.715}Co_{0.285}As_2$, for $EuFe_{1.715}Co_{0.285}As_2$ its Tc shows up at 24 K [21], as displayed in Fig.2c. The real part of ac susceptibility data for $EuFe_2As_2$, $EuFe_2As_{1.4}P_{0.6}$ sample is very consistent with dc results. However, the imaginary parts of ac susceptibility for the undoped and P-doped samples are found to be quite different. For comparison, we plotted ac susceptibility data of Co-doped sample in the same graph. We note that the imaginary part of the susceptibility is featureless for the undoped sample, but goes up for the P-doped and Co-doped samples. This increase of imaginary part below Tc for P-doped and Co-doped samples is clear indication of superconductivity together with the fact that they both exhibit zero resistivity, although the existing ferromagnetism lifts the susceptibility greatly hence no superconducting diamagnetism is directly detectable.

To obtain the evidence for real valence changes induced by internal and external pressure, XAS experiments on $L_3$-Eu absorption edge were carried out at SSRF. Fig.3a shows energy dependence of normalized XAS data collected at room temperature and at ambient pressure for the $EuFe_2As_2$, $EuFe_2As_{1.4}P_{0.6}$ and $EuFe_{1.715}Co_{0.285}As_2$ samples. The main peak and the satellite peak in the figure are associated to the divalent Eu and the trivalent Eu, respectively. We note that the intensity of the main peak goes downward and the intensity of the satellite goes upward simultaneously is only seen in the P-doped sample. The clear spectra weight transfer implies that the valence state of Eu ions in $EuFe_2As_{1.4}P_{0.6}$ is transferred from a divalent magnetic ($4f^7$, $J=7/2$) state into a trivalent nonmagnetic ($4f^6$, $J=0$) state. The valence transition should be caused by internal pressure. We estimated mean valence of Eu ions in $EuFe_2As_{1.4}P_{0.6}$ sample using widely used formula: $v=2+I^{Eu3+}/(I^{Eu2+}+I^{Eu3+})$, where $I^{Eu2+}$ and $I^{Eu3+}$ are amplitudes of peaks corresponding to $Eu^{2+}$ and $Eu^{3+}$ on XAS spectrum [22-23], to be about 2.32. Based on XAS data obtained, it is found that the internal pressure-induced valence transition from $Eu^{2+}$ to $Eu^{+2.32}$ provide additional 32% charges into FeAs layer of single crystal $EuFe_2As_{1.4}P_{0.6}$, comparable with the portion of charges (~29%) provided by $Co^{3+}$ in the optimal doped $EuFe_{1.715}Co_{0.285}As_2$. Our experimental results suggest that superconductivity in $EuFe_2As_{1.4}P_{0.6}$ is tightly related to Eu's valence transition.

High pressure XAS experiments on $EuFe_2As_2$ sample were also performed to get a deeper insight into the connection between valence change of Eu ions and pressure-induced superconducting transition in $EuFe_2As_2$ sample. The pressure dependence of $L_3$-Eu absorption spectra was plotted in Fig. 3(b). An obvious increase in intensity of the satellite peak which is associated with $Eu^{3+}$ configuration reflects individual response to high pressure of Eu ions exposed to different local environments. The valence transition of Eu ions begins at 2.1 GPa where Eu's mean valence is about 2.27. With increasing pressure, the mean valence of Eu ions is enhanced up to 2.33 and saturated mostly at pressure above 9.8 GPa, as displayed in inset of the Fig.3(b). The observed results demonstrate that physical (external) pressure can also drive Eu's valence transition from the divalent to the trivalent.

Physical pressure-induced superconductivity in $EuFe_2As_2$ has been found under high pressure [15-16]. The Tc shows up at 2.6-2.8 GPa where the mean valence of Eu ions is about 2.29-2.31, fairly consistent with the value (2.32) of P-doped sample. From above results, it can be concluded that the valence transition of Eu ions from the divalent to the trivalent is a physical origin of superconducting transition for compressed $EuFe_2As_2$ and $EuFe_2As_{1.4}P_{0.6}$, in which Eu layer provide additional charges to FeAs layer under pressure, as an alternative way of chemical doping. Whether the valence transition induced by chemical and physical pressure appears in other parent compounds of pnictide or other systems is an open issue and deserves further experimental and theoretical effort.

Acknowledgements

We sincerely thank X. Gao of SSRF for XAS experiment help. Authors wish to thank the National Science Foundation of China for its support of this research through Grant No. 10874230 and 10874211. This work was also supported by projects of 973 and Chinese Academy of Sciences. We acknowledge the support from EU under the project CoMePhS.

**Figure captions**

Figure 1. X-ray diffraction patterns of $EuFe_2As_2$ and $EuFe_2As_{1.4}P_{0.6}$ single crystals measured at room temperature with radiation of Cu K$\alpha$.

Figure 2. Temperature dependence of electronic resistance (a), dc susceptibility measured under magnetic field of 1000 Oe (b), real part and imaginary part of ac susceptibility measured at 100 Hz and 3 Oe (c) for $EuFe_2As_2$, $EuFe_2As_{1.4}P_{0.6}$ and $EuFe_{1.715}Co_{0.285}As_2$ single crystals.

Figure 3. X-ray absorption spectra data obtained at room temperature for three samples (a) and at high pressure for $EuFe_2As_2$ sample (b). Inset of figure 3(b) is an expanded plot of pressure dependence of Eu's mean valence in $EuFe_2As_2$ single crystal.

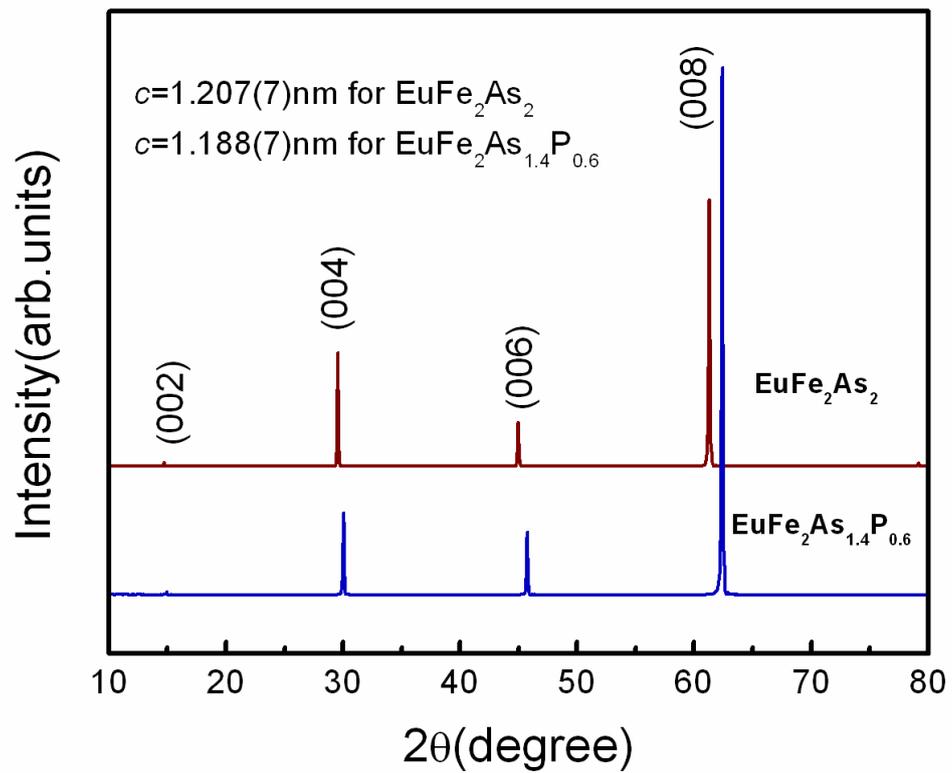

Fig. 1 Sun et al

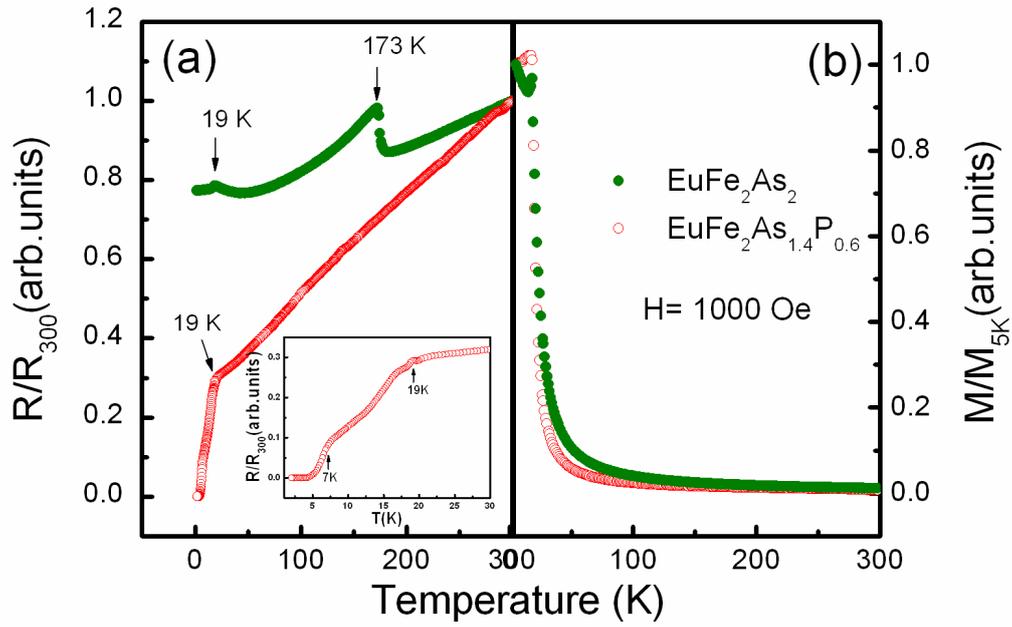

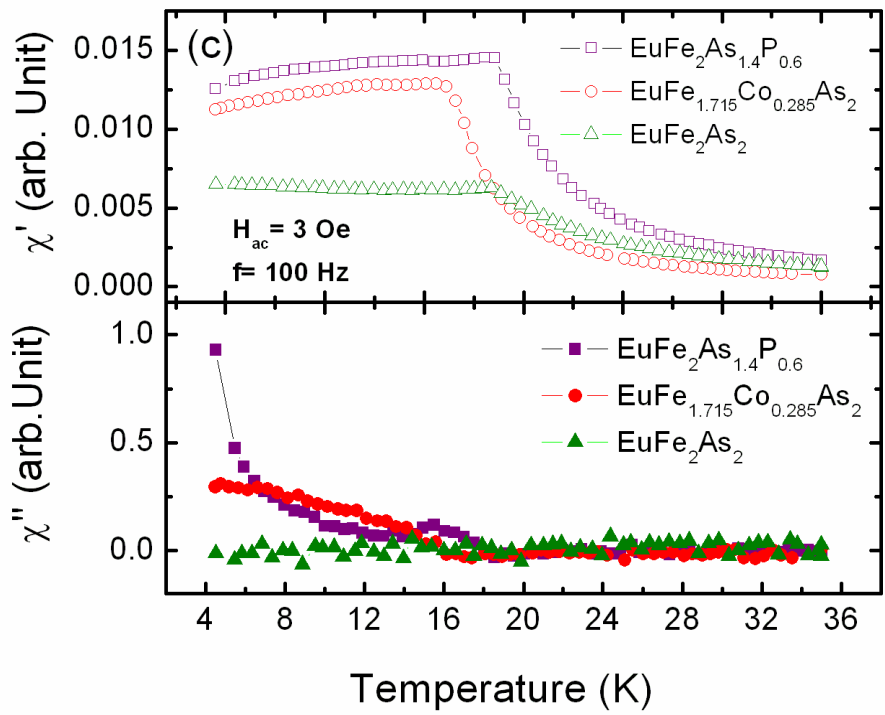

Fig. 2 Sun et al

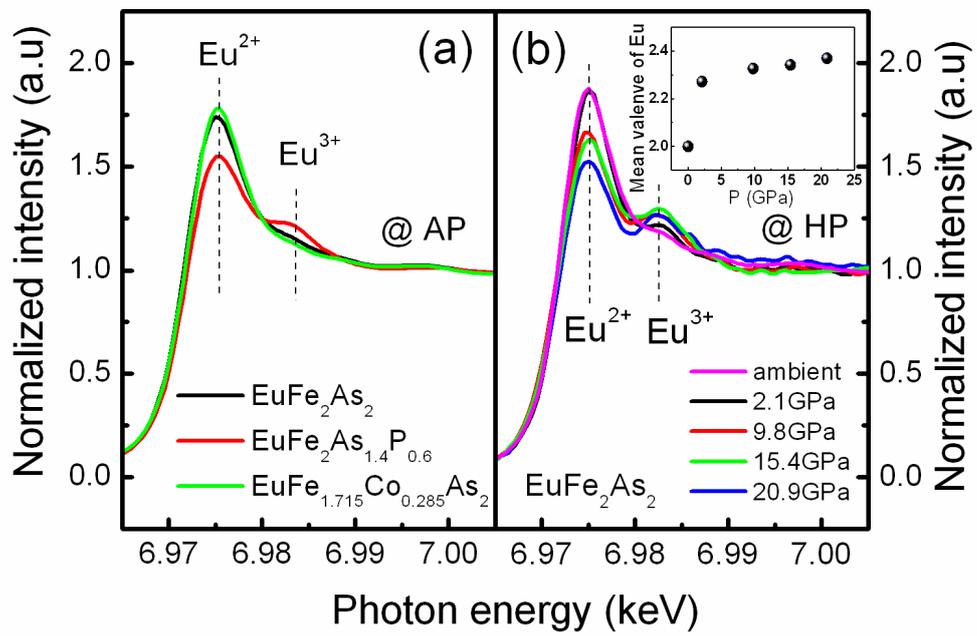

Fig.3 Sun et al